\documentclass[11pt,twoside]{article}
\usepackage{asp2004}
\usepackage{psfig}
\usepackage{epsf}
\usepackage{graphics}
\usepackage{lscape}
\begin{document}

\title{Irradiation-driven Mass Transfer Cycles in Compact Binaries}

\author{Andreas B\"uning and Hans Ritter}
\affil{Max-Planck-Institut f\"ur Astrophysik, 
       Karl-Schwarzschild-Str. 1, 
       D-85741 Garching bei M\"unchen}

\begin{abstract}
We elaborate on the analytical model of Ritter, Zhang, \& Kolb (2000)
which describes the basic physics of irradiation-driven mass transfer 
cycles in semi-detached compact binary systems. In particular, we take
into account a contribution to the thermal relaxation of the donor
star which is unrelated to irradiation and which was neglected in
previous studies. We present results of simulations of the evolution
of compact binaries undergoing mass transfer cycles, in particular
also of systems with a nuclear evolved donor star. These computations
have been carried out with a stellar evolution code which computes
mass transfer implicitly and models irradiation of the donor star in
a point source approximation, thereby allowing for much more
realistic simulations than were hitherto possible. We find that
low-mass X-ray binaries (LMXBs) and cataclysmic variables (CVs) with
orbital periods $\la 6$hr can undergo mass transfer cycles only for
low angular momentum loss rates. CVs containing a giant donor or one
near the terminal age main sequence are more stable than previously
thought, but can possibly also undergo mass transfer cycles. 

\end{abstract}

\section{Introduction}

The possible importance of irradiating the donor star of a
semi-detached compact binary by accretion luminosity for its long-term
evolution has first been pointed out by Podsiadlowski (1991).
Subsequently, the stability of mass transfer with irradiation feedback
has been studied in some detail by King et al. (1996, 1997, hereafter
KFKR96 and KFKR97), and by Ritter, Zhang, \& Kolb (2000, hereafter
RZK00). In KFKR96 and KFKR97 it was also shown that mass transfer in
cataclysmic variables (CVs) and low-mass X-ray binaries (LMXBs) can
become unstable against irradiation feedback, and that in case of
instability mass transfer proceeds in cycles in which episodes of
irradiation-driven mass transfer alternate with low-states during
which mass transfer is essentially shut off. Evolutionary calculations
of mass transfer cycles which were based on a homology model for the
donor and a simplified irradiation model, have been presented in KFKR97
and RZK00. Here we elaborate on the work done in previous studies in
two respects: First, we have developed a binary evolution code
which allows us to simulate the evolution of compact binaries with
mass transfer cycles more realistically than was possible in
previous studies, and second, we elaborate on the stability analysis of
KFKR96, KFKR97 and RZK00, thereby taking into account a contribution
to the thermal relaxation of the donor star which is unrelated to
irradiation and which has been neglected in previous studies. 

\section{Input Physics and Model Assumptions for the Numerical
Calculations}

In the following we are briefly listing the main model assumptions and
details of the input physics adopted for our numerical
calculations. For details the reader is referred to B\"uning \&
Ritter (2004), hereafter BR04. 

\subsection{The Stellar Evolution Code}
Basically, we use the 1D stellar evolution code described by Schlattl,
Weiss, \& Ludwig (1997) and Schlattl (1999). For calculating binary
evolution, in particular for determining the mass transfer rate
essentially free of numerical noise, considerable refinements in the
calculation of the equation of state and the opa\-cities were necessary.
A detailed description of what has been done and how, is given in BR04. 

\subsection{Computing Mass Transfer}

The mass loss rate from the donor star $-\dot M_2$ is computed
following Ritter (1988), i.e.\ from an explicit relation of the form 
\begin{equation}
-\dot M_2 = \dot M_0 \;e^{-\frac{R_{\rm R,2}-R_2}{H_{\rm P}}}\,
\end{equation}
where $R_2$ and $R_{\rm R,2}$ are respectively the radius of the donor
star and the corresponding critical Roche radius, $H_{\rm P}$ the
effective photospheric pressure scale height of the donor, and 
$\dot M_0\geq 0$ a slowly varying function of the donor's mass,
radius, and photospheric parameters. For our numerical computations
Eq.~(1) is formulated  as an outer boundary condition for the donor
star and is thus solved implicitly with the stellar structure
equations. The numerical setup is similar to what has been used by
Benvenuto \& de Vito (2003) and is described in detail in B\"uning
(2003, hereafter B03).  

\subsection{Irradiation Physics}

If the unperturbed donor star has a deep outer convective envelope,
i.e.\ if it is a cool star, irradiation, if not too strong, can be
treated as a local problem (for a detailed justification see RZK00).
Therefore, the problem reduces to specifying the intrinsic flux
$F_{\rm int}$, i.e.\ the true energy loss of the donor per unit
surface area and unit time as a function of the component of the
irradiating flux perpendicular to the stellar surface $F_{\rm irr}$.
For our numerical computations we use for $F_{\rm int}(F_{\rm irr})$
results tabulated by Hameury \& Ritter (1997) and additional data
kindly computed by Hameury (private communication) at our request.  

\subsection{Irradiation Model}

The irradiation model links the momentary accretion luminosity
$L_{\rm accr}$ (liberated by accretion onto the compact star) to  the
irradiating flux seen by a surface element of the irradiated donor
star. For our numerical computations we adopt the so-called point
source model which assumes that the spherical donor (of radius $R_2
\approx R_{\rm R,2}$) is illuminated by a point source with luminosity 
$L_{\rm accr}$ at the position of the accretor, i.e.\ at the orbital
distance $a$. Because the accretion luminosity is not necessarily
radiated isotropically or staedily from the accretor, and because only
a part of the irradiating flux is absorbed below the donor's
photosphere, the link between $F_{\rm irr}$ and $L_{\rm accr}$
\begin{equation}
F_{\rm irr} = \alpha \frac{L_{\rm accr}}{4 \pi a^2} h(\theta) 
            =: \left< F_{\rm irr} \right> \, h(\theta)
\end{equation} 
involves a dimensionless and a priori unknown efficiency parameter
$\alpha < 1$ which is the main unknown quantity in the irradiation
problem. In the framework of the point source model the flux
$F_{\rm irr}$ depends also on the substellar latitude $\theta$ of the
irradiated surface element. This is taken into account by the function
$h(\theta)$ in Eq.~(2) (for details see e.g.\ RZK00 or BR04). 

\section{Stability Analysis}

The stability of mass transfer in the presence of irradiation has been
studied previously in some detail by KFKR96, KFKR97 and RZK00. Because
some of the numerical results which we have obtained with the
above-described stellar evolution program were at variance with
results of these stability analyses, we have carried out a more refined
stability analysis and found that an additional term in the stability
criterion which becomes particularly important for giant donors had
been ignored in ealier considerations. Space limitations do not allow
us to repeat our stability analysis in detail. For this we refer the
reader to BR04. Rather we wish to give here only the main result. The
criterion for the stability of mass transfer can be written as
follows:  
\begin{equation}
\frac{ds}{d\ln{-\dot{M_2}}} < \frac{{\tau}_{\rm ce}}{{{\tau}^{'}}_{\rm d}}
                            + \frac{H_{\rm P}}{R_2}\, \delta 
\end{equation}
Here $s$ is the effective fraction of the donor surface through which
energy outflow from its interior is totally blocked by irradiation
(for a detailed definition of $s$ see BR04), $\tau_{\rm ce}$ is the
thermal time scale of the donor's convective envelope,
${{\tau}^{'}}_{\rm d}$ the time scale on which mass transfer is driven,
i.e.\  
\begin{equation}
\frac{1}{\tau^{'}}_{\rm d} = {\left(\frac{\partial \ln R_2}{\partial t}\right)}_{\rm nuc}
                           + {\left(\frac{\partial \ln R_2}{\partial
                           t}\right)}_{\rm th}
                           -\, 2 \, \frac{\partial \ln J}{\partial t}\,
\end{equation}
is the sum of the driving terms resulting from nuclear evolution and
thermal relaxation of the secondary, and from systemic loss of orbital
angular momentum $J$. In addition, if homology is used for describing
the structure of the secondary, $\delta$ in (3) can be approximated as  
\begin{equation} 
\delta = 4\,(1 - s) {\left(\frac{R_2}{R_{\rm 2,e}}\right)}^3 
       + (n + 1) {\left(\frac{R_2}{R_{\rm2, e}}\right)}^{-(n+2)} 
       \sim (n + 5),
\end{equation}
where $R_{\rm 2,e}$ is the thermal equilibrium radius of the
irradiated donor (under stationary irradiation) and $n=(\partial
\ln {\varepsilon}_{\rm nuc}/\partial \ln T)$ is the temperature
exponent of the nuclear energy generation rate ${\varepsilon}_
{\rm nuc}$. In the case of low-mass main sequence donors which burn
hydrogen via the pp-chain $n \approx 5$. In the case of giant donors
one has to set $n = -3$ for self-consistency. 

The new result is the last term in Eq.~(3). It derives from the
previously neglected fact that the thermal relaxation term, i.e.
$(\partial R_2/\partial t)_{\rm th}$, not only has a non-vanishing 
derivative with respect to $\Delta R = R_2 - R_{\rm R,2}$ but also
with respect to $\Delta R_{\rm e} = R_2 - R_{\rm 2,e}$. Since the 
last term of (3) is always positive it stabilizes mass transfer. Its
consequences are first that mass transfer with irradiation feedback
is always stable for very small driving rates, i.e.\ very small mass
transfer rates, and second that it is important for giant donors where
$H_{\rm P}/R_2$ is much larger than for main sequence stars.
Therefore,  contrary to earlier results obtained by KFKR97,
mass transfer from giant donors in CVs is much more stable than
previously thought. 

\section{Results}

Here we are mainly interested in answering the question whether 
irradiation-driven mass transfer cycles can occur in CVs or LMXBs. 
A necessary condition for such mass transfer cycles to be possible is
violation of the stability criterion (3). As can be seen from (3),
this is possible only if both terms on the right-hand side of (3) are
sufficiently small. It follows then immediately that mass transfer
cycles are most likely to occur if the donor star has a relatively
shallow convective envelope (i.e.\ a small $\tau_{\rm ce}$), and/or
mass transfer is driven on a long timescale ${\tau^{'}}_{\rm d}$, and
if the relative photospheric scale height, i.e.\ $H_{\rm P}/R_2$,
is small. On the other hand, the left-hand side of (3)
is itself also a function of the mass transfer rate (via the
irradiating flux) which vanishes for very small and very large
mass transfer rates and which attains a maximum value of $\sim 0.1$
for irradiating fluxes $1 \la \left< F_{\rm irr}\right>/F_0 \la 10$,
where $F_0$ is the unperturbed flux of the donor. Therefore, for a
given donor star (i.e.\ given values of $\tau_{\rm ce}$ and
$H_{\rm P}/R_2$)  there is always only a restricted range of mass
transfer rates, which also depends on the efficiency factor $\alpha$
defined in (2), for which mass transfer cycles can occur. Because
of (2) the left-hand side of (3) vanishes for both very small and very
large values of $\alpha$. For physical reasons we can exclude values
$\alpha \ga 1$. But apart from this upper limit $\alpha$ has to be
treated as a free parameter. 

From what has just been explained it has probably become clear that
irradiation--driven mass transfer cycles can occur only under special
conditions. Given the binary parameters, the rate of systemic angular
momentum loss $-\dot J$, and an adopted value of $\alpha$ one can
explore the range of instability by using (3) and an analytical
approximation for $\delta$, e.g.\ the one given in (5). In this way one
can narrow down the parameter space of interest before going to
extensive (and expensive) numerical calculations. For more details on
this point we refer the reader to BR04. 

\begin{figure}[ht!]
\plotone{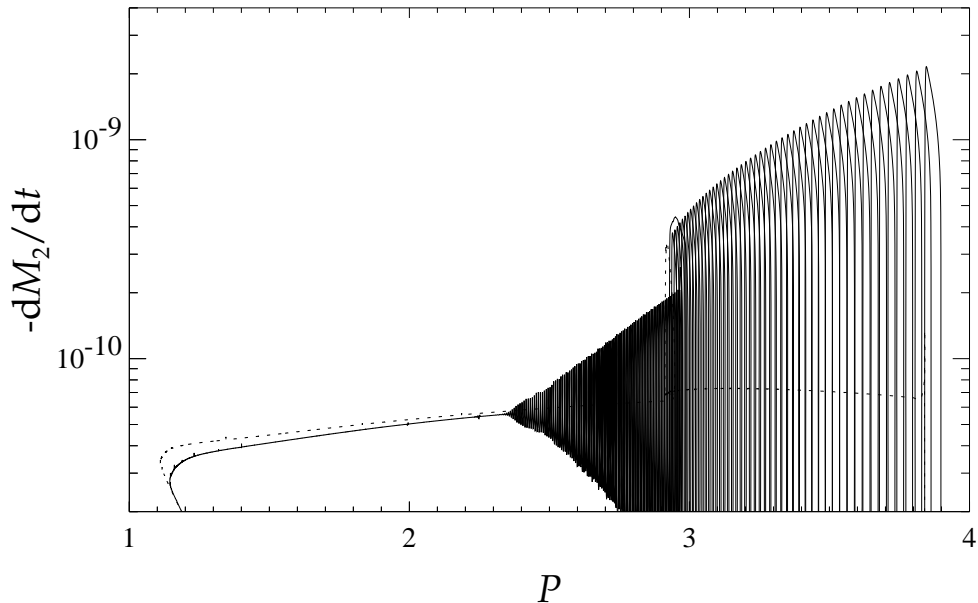}
\caption{Mass transfer rate (in $M_{\sun}$yr$^{-1}$) of a CV undergoing
         irradiation-driven mass transfer cycles (full line) as a
         function of its orbital period $P$ (in hours). The
         corresponding evolution without irradiation feedback is shown
         as a dashed line. For the parameters underlying these
         calculations see text.}

\end{figure}

As an example of such a numerical calculation we show in Fig.~1 the
variation of the mass transfer rate (in $M_{\sun}$yr$^{-1}$) as a
function of orbital period $P$ (in hours) of a CV undergoing mass
transfer cycles (full line). For this calculation we have
assumed the following parameters: The systemic loss of orbital angular
momentum is due to gravitational radiation only, i.e.\ $\dot J = {\dot
J}_{\rm GR}$. The donor star is a standard Pop.~I low-mass main
sequence star with an initial mass $M_2 = 0.5 M_{\sun}$, an age of
$10^{10}$yr and a central hydrogen mass fraction of $X_{\rm c} \approx
0.62$. In this case $H_{\rm P}/R_2 \approx 10^{-4}$. The primary is a
white dwarf with a mass of $M_1 = 0.8 M_{\sun}$ and a radius of $R_1 =
0.010 R_{\sun}$. The transferred mass is assumed to eventually leave
the binary system (e.g.\ via nova explosions) with a dimensionless
orbital angular momentum $\partial \ln J_{\rm orb}/\partial  \ln (M_1
+ M_2) = M_2/M_1$. The irradiation efficiency parameter is $\alpha =
0.3$. The dotted line shows the corresponding evolution without
irradiation feedback. 

What this example shows, is first how numerically accurate the stellar
evolution code described in Sect.~2 can follow such an evolution.
Second, we see that irradiation-driven mass transfer cycles can occur
in CVs for not unrealistic values of $\alpha$ provided that the
braking rate is small. In a corresponding evolution in which the
absolute value of the angular momentum loss rate is much higher, e.g.\
according to the Verbunt \& Zwaan (1981) prescription, no mass
transfer cycles would occur. We note that the step in the amplitude of
the mass transfer cycles at an orbital period of $\sim 3$ hr is due to
the secondary becoming fully convective. Third, because the thermal
time scale of the fully convective donor star increases with
decreasing mass, mass transfer eventually becomes stable at $P \approx
2.5$ hr. 

Based on numerous numerical simulations (detailed in B03 or BR04) 
and on the more general considerations outlined above we can summarize
our results as follows: 

\begin{enumerate}
\item In agreement with earlier results we find that CVs which contain
       a main sequence donor and in which the driving rate above the
       period gap is as high as required for explaining the period gap
       in the framework of the model of disrupted magnetic braking
       (see e.g.\ Spruit \& Ritter 1983) are stable against
       irradiation feedback except for the most massive donor stars
       $M_2 \sim 1 M_{\sun}$. On the other hand, mass transfer cycles
       can occur in short-period CVs if the driving rate is small,
       i.e.\  no larger than a few times the gravitational braking rate.

\item CVs containing a donor star which is near the terminal age main
      sequence turn out to be more stable than has been anticipated
      based on simple homology arguments.   

\item CVs containing an extended giant donor with nuclear
      timescale--driven mass transfer are less likely to undergo mass
      transfer cycles than has been anticipated based on the results
      of KFKR97. The main reason for this discrepancy is the second
      term on the right-hand side of the stability criterion (3),
      i.e.\ the relatively large value of $H_{\rm P}/R_2$ associated
      with such stars. If mass transfer cycles do occur they are
      characterized by comparatively very short phases with high,  
      irradiation-driven mass trasnsfer rates which are followed by
      extended periods without mass transfer during which the system
      is essentially detached and reattachment is reached only on the
      nuclear time scale of the giant. 

\item Adopting the point source irradiation model which takes into
      account irradiation of surface elements near the terminator of
      the donor we find that possibly also LMXBs can undergo mass
      transfer cycles. Regarding the braking rate which is necessary
      to drive cycles, basically the same restrictions apply as for
      short--period CVs. We confirm also that LMXBs containing a giant
      donor can undergo cyles. 

\item Mass transfer cycles in CVs do occur only if $0.1 \la \alpha \la
      1$ whereas in the case of LMXBs cycles do not occur if $\alpha
      \ga 0.1$. 

\item For systems containing an unevolved main sequence or a giant
      donor the results of our numerical computations and the
      predictions from the analytic model for the stability boundaries
      are in reasonable agreement.
\end{enumerate}

\end{document}